%% file: main.tex
\documentclass[wcp]{jmlr}

\usepackage{array}
\usepackage{enumitem}
\usepackage{multirow}
\usepackage[table]{xcolor}
\usepackage[most]{tcolorbox}
\usepackage{booktabs}

\usepackage{lineno}
\makeatletter
\let\Ginclude@graphics\@org@Ginclude@graphics
\makeatother

\input{sections/macros}

\jmlrvolume{353,}
\jmlryear{2026}
\jmlrworkshop{ACML 2026}

\title[\method: Minimum-Necessary Disclosure]{\method: Enforcing Minimum-Necessary Disclosure in Tool-Calling LLM Agents}

\author{\Name{Ying Fu} \Email{fuying@stu.xidian.edu.cn}\\
  \Name{Yulong Fu} \Email{ylfu@xidian.edu.cn}\\
  \addr Xidian University}

\editors{Andy Song, Bo Han and Sarah Erfani}

\begin{document}

\makeatletter
\let\@jmlrpages\@empty
\makeatother

\maketitle

\input{sections/abstract}


\input{sections/introduction}
\input{sections/related_work}
\input{sections/method}
\input{sections/experiments}
\input{sections/limitations}
\input{sections/conclusion}

\acks{This work was supported by the 2026 Open Research Fund of the Key Laboratory of Software Integration Application and Test Verification, Ministry of Industry and Information Technology (No. RFT20260005), the Tarim University President Fund (TDZKCX202509), and the Tencent Cloud Computing (Beijing) Co., Ltd. Grant No. HX01202203011.}

\begingroup
\small
\setlength{\bibsep}{0pt plus 0.3ex}
\bibliography{sections/references}
\endgroup

\end{document}

%% file: sections/macros.tex
\newcommand{\method}{ArguShield}
\newcommand{\dataset}{ArgPrivBench}
\newcommand{\keep}{\textsc{Keep}}
\newcommand{\hide}{\textsc{Hide}}
\newcommand{\ask}{\textsc{Ask}}
\newcommand{\apvr}{APVR}
\newcommand{\tsr}{TSR}
\newcommand{\tsraa}{\ensuremath{\mathrm{TSR}_{\text{approve-all}}}}
\newcommand{\askr}{ASKR}

%% file: sections/abstract.tex
\begin{abstract}
    Large language model agents increasingly use third-party tools to complete user tasks. Even without malicious intent, they may fill broad or weakly constrained fields with private information that is unnecessary or inappropriate for the receiving tool. Such calls can satisfy the tool schema and complete the task, yet still violate minimum-necessary disclosure. We introduce \method{}, a mediation layer that intercepts proposed tool calls before execution. It evaluates each candidate value using three signals: Value Provenance, Task Relevance, and Disclosure Boundary. \method{} releases justified values, defers values that need user confirmation, and withholds those outside the permitted scope. We evaluate \method{} with four tool-calling models on \dataset{}, a benchmark of 120 tasks across Travel, Shopping, and Healthcare. Unprotected agents frequently over-disclose private information despite high task success. Prompt-only protection reduces unauthorized disclosure but substantially lowers task success. \method{} reduces unauthorized silent release to 9\% on Standard tasks and 6\% on High-risk tasks while preserving recoverable task success under simulated user confirmation.
    \end{abstract}

%% file: sections/introduction.tex
\section{Introduction}
\label{sec:introduction}

Recent AI agents increasingly complete user tasks through third-party tools. Locally deployable agents with persistent memory and external application access, such as OpenClaw~\citep{wang2026openclawsecurity}, expand the boundary at which private context may leave the system: a value in memory, profile, or context becomes an external disclosure once copied into a tool call.

This disclosure is not always caused by malicious intent or incorrect tool choice. Third-party schemas may expose broad, optional, or weakly constrained fields; 36.9\% of 2,344 Action/API specifications in the OpenAI GPT ecosystem expose at least one weakly constrained free-text parameter such as \texttt{query}, \texttt{notes}, or \texttt{context} \citep{shayesteh-wilson-2026-conventional}. Such fields make over-disclosure easy to express as a schema-valid call that completes the task while still transmitting values the receiver does not need.

\begin{figure}[t]
\centering
\includegraphics[width=\linewidth]{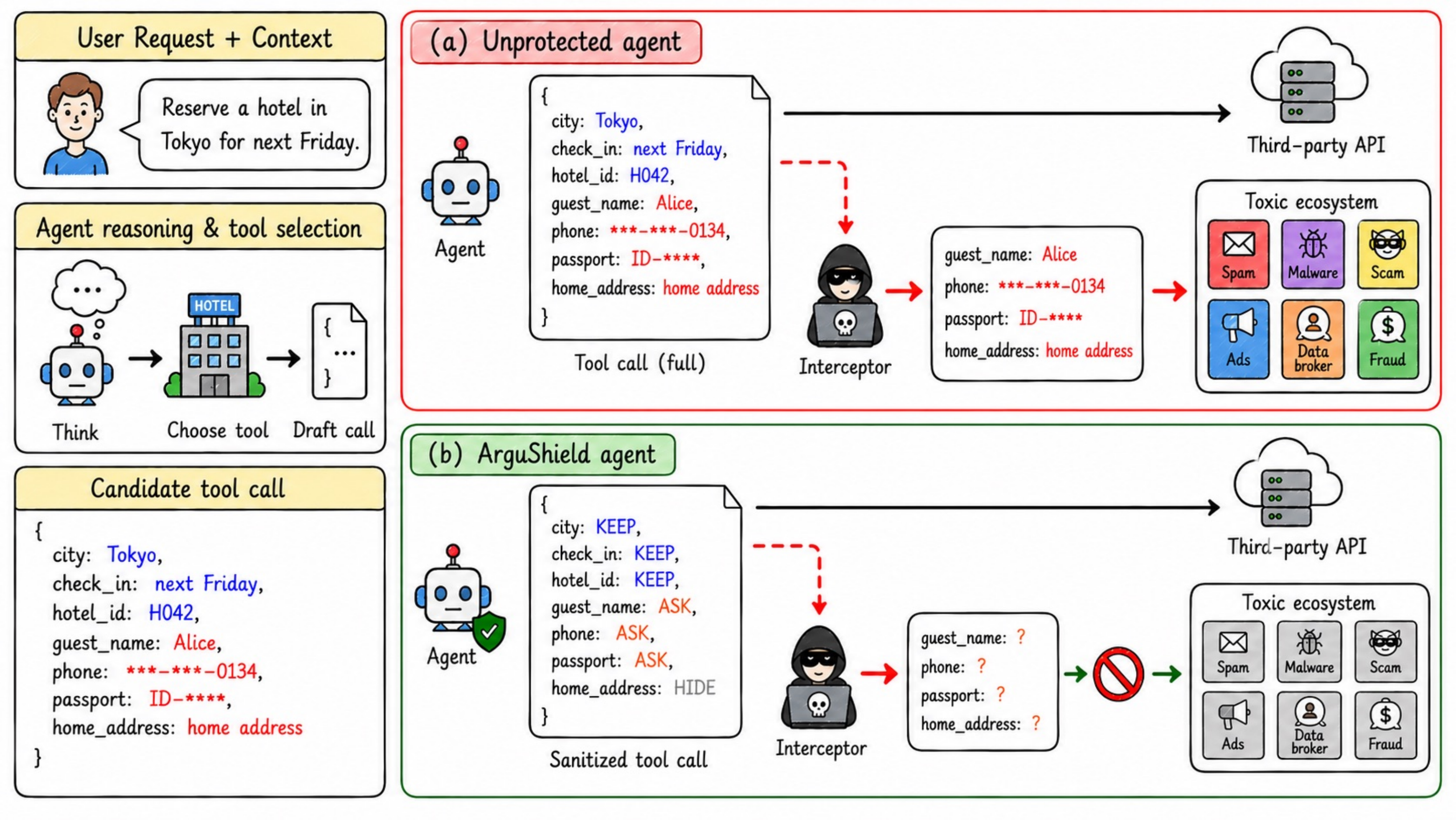}
\caption{Motivating example of minimum-necessary disclosure in third-party tool calling. An unprotected agent sends a full tool call that mixes operational task fields with profile-derived personal values. \method{} mediates the proposed invocation before execution: task-needed fields are released, confirmation-needed personal values are deferred, and values outside the disclosure boundary are withheld. \keep{}, \ask{}, and \hide{} denote value-level release decisions rather than literal argument values sent to the API.}
\label{fig:motivation}
\end{figure}

Figure~\ref{fig:motivation} illustrates this failure for hotel reservation: operational fields can be mixed with profile-derived name, phone, passport, and address values, some requiring confirmation and others falling outside the task disclosure boundary.

We study this as a violation of \emph{minimum-necessary disclosure}: a value should be released only when the current task and receiving tool justify it. We call such violations \emph{tool-argument over-disclosure}. Task completion and appropriate information flow are distinct; purpose-bound privacy evaluation shows private information can reach unnecessary sinks even when the task succeeds \citep{hu2026toolprivacybench}.

We introduce \method{}, a pre-execution release mediation layer that treats model-proposed invocations as drafts, evaluates each candidate value with \emph{Value Provenance}, \emph{Task Relevance}, and \emph{Disclosure Boundary}, and assigns \keep{}, \ask{}, or \hide{} before execution.

We evaluate on \dataset{}, a 120-task benchmark across Travel, Shopping, and Healthcare (20 Standard and 20 High-risk tasks per domain), with annotated release labels that separate task utility from disclosure necessity.

Across four tool-calling models, unprotected agents often succeed yet over-disclose; prompt-only protection reduces unauthorized disclosure at a substantial cost to task success. \method{} reduces unauthorized silent release on Standard and High-risk tasks while preserving recoverable task success under simulated user confirmation.

Our contributions are:
\begin{itemize}
\item We formulate minimum-necessary disclosure for third-party tool-calling agents and identify \emph{tool-argument over-disclosure} as a central failure mode.
\item We propose \method{}, a pre-execution mediation layer enforcing value-release decisions via Value Provenance, Task Relevance, and Disclosure Boundary.
\item We introduce \dataset{} (120 tasks, three domains) and show \method{} reduces unauthorized silent release while preserving recoverable task success under simulated user confirmation.
\end{itemize}

%% file: sections/related_work.tex
\section{Related Work}
\label{sec:related-work}

\paragraph{Tool-use and agent evaluation.}
Tool use has become a central interface for extending language models from text generation to external action. Early work studies reasoning with actions, self-supervised tool-use learning, and generalization across large API collections \citep{yao2023react,schick2023toolformer,qin2024toolllm}. Subsequent work addresses API-call generation and function-calling evaluation \citep{patil2024gorilla,patil2025bfcl}, as well as verified function-calling dataset construction \citep{liu2024apigen}. More recent agent environments move from isolated API calls to stateful interaction, multi-turn workflows, and tool-agent-user settings \citep{wang2024hammerbench,lu2025toolsandbox,yao2025taubench}. Broader web, app, desktop, and workplace benchmarks further expand the evaluation of agents in realistic interfaces \citep{trivedi2024appworld,zhou2024webarena,xie2024osworld}. These studies establish the importance of tool calling and action execution, but they primarily measure whether agents can complete tasks. Our work studies a complementary property: whether the values released to a third-party tool are minimally necessary for the current task and receiver.

\paragraph{Agentic disclosure and purpose-bound privacy.}
Privacy risks in tool-calling agents arise not only from model behavior, but also from the interfaces agents are asked to populate. Third-party tool schemas can expose broad, optional, or weakly constrained fields, leaving disclosure decisions to the agent at runtime \citep{shayesteh-wilson-2026-conventional}. AgentDAM evaluates privacy leakage in autonomous web agents, while SPILLage studies agentic oversharing on the web \citep{zharmagambetov2025agentdam,roh2026spillage}. Prompt-injection attacks can induce leakage of personal data observed by agents \citep{alizadeh2025promptleakage}, whereas PrivacyAsst proposes a privacy-protection method for tool-using agents \citep{zhang2024privacyasst}. Other work frames privacy in terms of contextual integrity, purpose, recipient, and situation rather than sensitivity alone \citep{bagdasarian2024airgapagent,mireshghallah2024confaide,li2025privaci}. ToolPrivacyBench formalizes a closely related purpose-bound privacy problem: successful tool execution does not imply appropriate disclosure, because private information may be routed to tools or downstream sinks that do not require it \citep{hu2026toolprivacybench}. \method{} targets this failure at the point where it becomes executable: before a proposed invocation releases candidate values to a third-party API.

\paragraph{Tool-boundary mediation and enforcement granularity.}
Existing defenses differ in their intervention points, policy representations, and enforcement granularity. \citet{kan2023privacytextsanitization} sanitize sensitive inputs before remote model access, whereas \citet{wen2026contextualized} generate context-specific privacy guidance during agent execution. PrivacyLens~\citep{shao2024privacylens} provides a complementary framework for evaluating privacy violations in agents' actions. Mobile-agent systems use visual interface context to plan and execute actions \citep{wang2024mobileagent,zhang2025appagent}. Privacy-focused evaluation further reveals limitations in smartphone agents' recognition of privacy-sensitive situations \citep{lin2026sapa}. For tool-calling agents, however, the receiving tool may observe private values before any final response exists. Tool-risk and agent-security studies show that tool use can cause privacy leakage, unsafe execution, prompt-injection vulnerabilities, speculative-call exposure, or multi-tool leakage channels \citep{ruan2024toolemu,debenedetti2024agentdojo,mohammadi2026ghost,qiao2025topr}. System-level defenses such as CaMeL and FIDES separate control/data flow or enforce information-flow policies to secure agents against untrusted inputs \citep{debenedetti2025camel,costa2025fides}. PACT makes a related granularity observation: risks can arise when untrusted information determines an authority-bearing argument, so provenance must be tracked below the whole-invocation level \citep{fan2026pact}. \method{} applies this argument-level perspective to privacy at the tool-execution boundary.

%% file: sections/method.tex
\section{\method{}: Enforcing Minimum-Necessary Disclosure at the Tool Boundary}
\label{sec:method}

\method{} is a pre-execution release mediation layer for tool-calling agents. Rather than executing a model-proposed tool call directly, \method{} treats it as an executable draft whose values must be justified before crossing the third-party tool boundary. The enforcement goal is \emph{minimum-necessary disclosure}: a value is released only when it is grounded in an appropriate source, warranted by the current task, and appropriate for the receiving tool.

\begin{figure}[t]
\centering
\includegraphics[width=\linewidth]{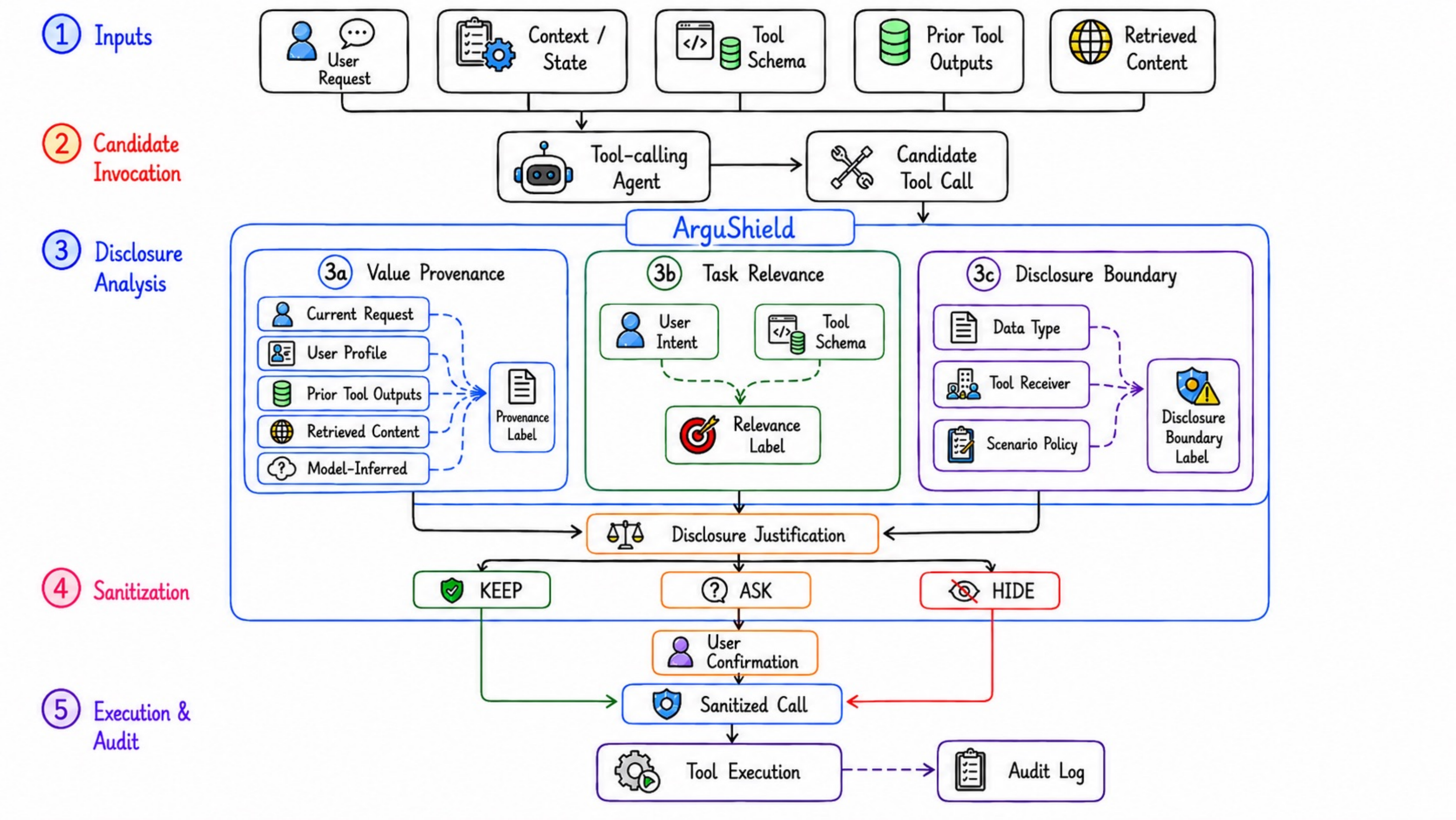}
\caption{Overview of \method{}. A model-proposed tool call is mediated before execution using Value Provenance, Task Relevance, and Disclosure Boundary, producing a sanitized call, confirmation queue, and audit log. The \hide{} branch denotes removal from the outgoing invocation, not transmission to the tool.}
\label{fig:system-overview}
\end{figure}

A schema-valid invocation may mix values with different disclosure status. Some values are required for task completion, some may be useful but require user confirmation, and others are outside the permitted disclosure boundary. \method{} therefore preserves the tool-level intent whenever possible while mediating which candidate values are allowed to leave the agent. Figure~\ref{fig:system-overview} summarizes the pipeline.

\subsection{Problem Formulation}
\label{sec:problem-formulation}

For a task, the agent observes the user request $x$, context/state $C$, target tool schema $S_t$, prior tool outputs $O$, and retrieved content $R$. The model proposes a structured invocation
\[
c=(t,A),
\]
where $t$ is the receiving tool and $A$ is the argument object. \method{} outputs a sanitized invocation $\tilde{c}=(t,\tilde{A})$, a confirmation queue $Q$, and an audit log $G$. Only $\tilde{c}$ is released to the tool. Values in $Q$ remain local unless the user confirms them, and blocked values are not transmitted.

The argument object is flattened into leaf-level candidate values
\[
\mathcal{L}(A)=\{\ell_i=(p_i,v_i)\}_{i=1}^{n},
\]
where $p_i$ is the argument path and $v_i$ is the proposed value. Each candidate receives one action
\[
a_i\in\{\keep{},\ask{},\hide{}\}.
\]
\keep{} authorizes immediate release, \ask{} defers release pending confirmation, and \hide{} removes the value from the outgoing call. This leaf-level formulation is important because a single invocation can contain both task-needed and over-disclosed values.

\subsection{Disclosure Justification}
\label{sec:disclosure-justification}

For each candidate value, \method{} constructs a disclosure justification record from three evidence signals. \emph{Value Provenance} asks where the value comes from. Values explicitly provided in the current request are treated differently from values copied from user profile, prior tool outputs, retrieved content, or model inference. This distinction prevents the agent from treating contextual availability as permission to disclose. For example, a home address stored in profile may be available to the agent, but it should not be silently copied into a hotel reservation unless the task and receiver justify that release.

\emph{Task Relevance} asks whether the current task warrants the value. A value can be required for task success, useful but not strictly necessary, unsupported by the request, or conflicting with the task state. This signal prevents optional schema fields from becoming automatic disclosure channels. A field such as \texttt{notes}, \texttt{context}, or \texttt{special\_instructions} may be schema-valid, but schema compatibility alone does not make every available personal value task-relevant.

\emph{Disclosure Boundary} asks whether the receiving tool is an appropriate recipient under the current purpose. Some values are justified for silent release, some may support task completion but require confirmation, and others should be withheld for the current receiver. This receiver-aware signal captures the fact that sensitivity alone is insufficient: the same phone number, address, or medical note can be appropriate for one tool, confirmation-needed for another, and unjustified for a different third-party service.

At inference time, \method{} does not use gold release labels. It estimates the three evidence signals from the runtime state and then applies a fixed release monitor. For Value Provenance, it performs normalized matching against the current request, profile fields, prior tool outputs, and retrieved content. Exact matches, normalized string matches, and typed matches for dates, names, addresses, identifiers, prices, and contact fields are assigned to their supporting source; unmatched or fabricated values are marked as inferred or unknown. For Task Relevance, it combines the user intent, tool schema, argument path, value type, and task-success criteria to determine whether the value is required, useful, unsupported, or conflicting. For Disclosure Boundary, it uses the receiver identity, tool purpose, argument path, value type, and authorization status to decide whether the value is permitted for silent release, requires confirmation, or is outside the receiver boundary. Ambiguous evidence is handled conservatively and cannot justify silent release.

\subsection{Release Policy and Execution}
\label{sec:release-policy}

After evidence extraction, \method{} applies a deterministic release monitor to each candidate value. The final action is fixed given the evidence record, making release behavior auditable and reproducible rather than delegated to another model judgment. The monitor follows a priority order. \hide{} has highest priority and is assigned when a value is unsupported, conflicting, inferred without reliable grounding, or outside the receiving tool's disclosure boundary. This rule blocks values that are schema-valid but not justified for release.

\keep{} is assigned only when all three disclosure conditions are satisfied: the value is source-supported, warranted by the task, and permitted for silent release to the current receiver. This covers operational arguments explicitly provided by the user for the current task or values otherwise grounded and necessary for execution. \ask{} is assigned when a value may support the task but lacks sufficient authorization for silent release, such as profile-derived or previously observed values whose transmission to the current receiver should be confirmed by the user. Thus, \ask{} is not a weaker form of \keep{}; it is a local withholding action that prevents the value from crossing the tool boundary until approval.

After assigning actions, \method{} reconstructs the outgoing invocation:
\[
\tilde{c}=(t,\tilde{A}),\qquad
Q=\{\ell_i \mid a_i=\ask{}\}.
\]
Values assigned \keep{} remain in $\tilde{A}$ and are sent to the tool. Values assigned \ask{} are removed from the outgoing call and placed in $Q$ until approved. Values assigned \hide{} are removed and are not offered for release. The audit log $G$ records each argument path, the resolved evidence labels, and the final action, enabling post-hoc inspection and error analysis.

\method{} is intended to complement, rather than replace, the underlying tool-calling model. The model still selects the tool and drafts the invocation, but release control is moved to the execution boundary where structured values can be checked before they leave the system. This separation allows task execution to proceed with the minimum set of justified values while keeping confirmation-needed and unjustified values local.

%% file: sections/experiments.tex
\section{Experiments}
\label{sec:experiments}

We evaluate whether tool-calling agents satisfy minimum-necessary disclosure when private context becomes executable tool input. Each instance provides a user request, visible context, and target tool schema. The model proposes a structured invocation, and we evaluate which values would be released to the third-party receiver. The evaluation is offline, but it preserves the failure studied in this paper: a call may be schema-valid and task-completing while still releasing values that the current task or receiver does not justify.

\subsection{Experimental Setup}
\label{sec:exp-setup}

\paragraph{Benchmark.}
\dataset{} contains 120 synthetic tasks across Travel, Shopping, and Healthcare, with 20 Standard and 20 High-risk tasks per domain. The domains cover common third-party tool interactions: Travel includes hotel booking, itinerary planning, and appointment scheduling; Shopping includes checkout, delivery, return, and coupon-related tools; Healthcare includes clinic booking, pharmacy, insurance, and patient-intake tools. Each task specifies a user intent, visible user/context state, target tool schema, task-specific success criteria, and a set of candidate values that may be copied into the invocation. Candidate values include operational fields needed by the tool, profile-derived values, prior-context values, and receiver-sensitive values. They are annotated as \keep{} when silent release is justified, \ask{} when the value may support completion but requires user confirmation, and \hide{} when it is outside the task need or disclosure boundary. Standard tasks contain everyday personal values and optional arguments where over-disclosure is plausible. High-risk tasks do not necessarily have higher violation frequency; instead, they involve stricter receiver, purpose, or value-type constraints, so an unjustified release is more severe.

\paragraph{Models, methods, and metrics.}
We evaluate DeepSeek-V4-Flash, GLM-5.1, Kimi-K2.7-Code, and Qwen3.7-Plus. We select these models because they support structured tool calls or schema-constrained argument generation, span four distinct model providers and families, and were available through commercial APIs during the evaluation period. All four models are evaluated with the same tasks, tool schemas, prompts, mediator configuration, and temperature of 0.0; only endpoint-specific tool-call formatting differs where necessary. No model is selected or excluded based on its performance with \method{}. \textsc{NoProtection} executes the generated call directly. \textsc{Prompt-Only Guard} adds a privacy instruction but does not inspect the structured call after generation. \method{} mediates the proposed call before execution. \apvr{} measures unauthorized silent release, namely gold \ask{}/\hide{} values released as \keep{}; it does not measure disclosure following simulated approval. TLR is the fraction of tasks with at least one unauthorized silent release. \tsr{} measures task success. In the main protection comparison, we report \tsraa{}, where task-required values routed to \ask{} are treated as recoverable after simulated approval; for \textsc{NoProtection}, the same column reports ordinary task success without confirmation. ADF1 is macro-F1 over \keep{}, \ask{}, and \hide{}. \askr{} measures the fraction of values routed to \ask{} on completed tasks, and is reported only as confirmation overhead. Values routed to \ask{} are not counted as released before confirmation.

\subsection{Over-Disclosure in Unprotected Calls}
\label{sec:over-disclosure}

Under direct execution, agents often select plausible tools and satisfy the user request, yet still copy optional profile, delivery, payment-adjacent, insurance, or medical-note values into the call (Table~\ref{tab:unprotected}). Averaged over all entries, unprotected execution yields 0.81 \apvr{} on Standard tasks, 0.77 on High-risk tasks, and 0.96 \tsr{} overall. The gap between high \tsr{} and high \apvr{} shows that task success alone does not imply appropriate information release.

\begin{table}[t]
\centering
\footnotesize
\renewcommand{\arraystretch}{1.18}
\setlength{\tabcolsep}{5.5pt}
\caption{Unprotected calls by domain and model.}
\label{tab:unprotected}
\begin{tabular}{llcccc}
\toprule
\textbf{Domain} & \textbf{Model}
& \multicolumn{2}{c}{\textbf{Standard}}
& \multicolumn{2}{c}{\textbf{High-risk}} \\
\cmidrule(lr){3-4}\cmidrule(lr){5-6}
& & \apvr & \tsr & \apvr & \tsr \\
\midrule
Travel
& DeepSeek-V4-Flash & 0.85 & 1.00 & 0.80 & 1.00 \\
& GLM-5.1           & 1.00 & 1.00 & 0.95 & 1.00 \\
& Kimi-K2.7-Code    & 0.80 & 1.00 & 0.85 & 1.00 \\
& Qwen3.7-Plus      & 0.80 & 1.00 & 0.60 & 1.00 \\
\midrule
Shopping
& DeepSeek-V4-Flash & 0.70 & 0.95 & 0.60 & 0.85 \\
& GLM-5.1           & 0.55 & 0.80 & 0.65 & 0.80 \\
& Kimi-K2.7-Code    & 0.85 & 0.85 & 0.95 & 0.95 \\
& Qwen3.7-Plus      & 0.90 & 0.95 & 0.85 & 0.95 \\
\midrule
Healthcare
& DeepSeek-V4-Flash & 0.70 & 1.00 & 0.60 & 1.00 \\
& GLM-5.1           & 1.00 & 1.00 & 0.80 & 1.00 \\
& Kimi-K2.7-Code    & 0.80 & 1.00 & 0.80 & 1.00 \\
& Qwen3.7-Plus      & 0.80 & 1.00 & 0.80 & 1.00 \\
\bottomrule
\end{tabular}
\end{table}

\subsection{Effectiveness of \method{}}
\label{sec:method-effectiveness}

Release mediation substantially reduces unauthorized silent release while preserving recoverable task success under simulated user confirmation (Table~\ref{tab:protection}). Averaged over four models, Standard \apvr{} drops from 0.84 under \textsc{NoProtection} to 0.09 with \method{}, and High-risk \apvr{} drops from 0.76 to 0.06. \tsraa{} remains 0.97 on both splits, and ADF1 improves consistently, indicating closer alignment with the intended \keep{}/\ask{}/\hide{} policy.

\begin{table}[t]
\centering
\footnotesize
\renewcommand{\arraystretch}{1.08}
\setlength{\tabcolsep}{2.0pt}
\caption{Protection results. \tsraa{} treats task-required \ask{} values as recoverable after simulated approval. For \textsc{NoProtection}, the same column reports ordinary task success without confirmation.}
\label{tab:protection}
\begin{tabular}{llcccccc}
\toprule
\textbf{Model} & \textbf{Method}
& \multicolumn{3}{c}{\textbf{Standard}}
& \multicolumn{3}{c}{\textbf{High-risk}} \\
\cmidrule(lr){3-5}\cmidrule(lr){6-8}
& & \apvr & \shortstack{TSR\\{\scriptsize (Approve-all)}} & ADF1 & \apvr & \shortstack{TSR\\{\scriptsize (Approve-all)}} & ADF1 \\
\midrule
DeepSeek-V4-Flash
& \textsc{NoProtection}      & 0.77 & 0.98 & 0.27 & 0.65 & 0.95 & 0.27 \\
& \textsc{Prompt-Only Guard} & 0.00 & 0.55 & 0.18 & 0.02 & 0.48 & 0.16 \\
& \method{}                  & 0.05 & 0.98 & 0.62 & 0.03 & 0.95 & 0.56 \\
\midrule
GLM-5.1
& \textsc{NoProtection}      & 0.88 & 0.97 & 0.26 & 0.83 & 0.93 & 0.26 \\
& \textsc{Prompt-Only Guard} & 0.07 & 0.75 & 0.25 & 0.05 & 0.65 & 0.22 \\
& \method{}                  & 0.15 & 0.97 & 0.61 & 0.09 & 0.93 & 0.60 \\
\midrule
Kimi-K2.7-Code
& \textsc{NoProtection}      & 0.83 & 0.93 & 0.25 & 0.85 & 0.98 & 0.26 \\
& \textsc{Prompt-Only Guard} & 0.05 & 0.82 & 0.27 & 0.08 & 0.78 & 0.26 \\
& \method{}                  & 0.12 & 0.93 & 0.59 & 0.08 & 0.98 & 0.63 \\
\midrule
Qwen3.7-Plus
& \textsc{NoProtection}      & 0.87 & 1.00 & 0.27 & 0.72 & 1.00 & 0.28 \\
& \textsc{Prompt-Only Guard} & 0.02 & 0.87 & 0.29 & 0.00 & 0.85 & 0.28 \\
& \method{}                  & 0.03 & 1.00 & 0.67 & 0.03 & 1.00 & 0.60 \\
\bottomrule
\end{tabular}
\end{table}

Prompt-only guidance achieves lower or equal \apvr{} across the evaluated settings, but substantially reduces task success and yields lower ADF1. In contrast, \method{} retains task-needed values and blocks or defers disclosures without sufficient task or receiver justification.

\subsection{External Validation Beyond \dataset{}}
\label{sec:external-validation}

We next test whether the fixed release policy transfers beyond the controlled \dataset{} setting. First, we collect tool schemas from 26 publicly available real-world MCP servers, including Shopify and GitHub, and construct privacy-sensitive scenarios around their original tool definitions. We execute the schemas through local stdio MCP servers and measure leakage from the arguments actually received after \texttt{tools/call}. Second, we evaluate 123 privacy-task instances from AgentDAM across Shopping, GitLab, and Reddit. Because AgentDAM represents web-agent actions rather than structured tool calls, we add an adapter only at the execution boundary to expose each proposed action as structured arguments; the original tasks, privacy semantics, environments, and success criteria remain unchanged. Neither evaluation retrains \method{} or changes its release policy.

\begin{table}[t]
\centering
\small
\renewcommand{\arraystretch}{1.12}
\setlength{\tabcolsep}{4.2pt}
\caption{External validation on independently developed MCP server schemas and AgentDAM privacy tasks. TLR is the fraction of tasks with at least one unauthorized silent release.}
\label{tab:external-validation}
\begin{tabular}{llrcccc}
\toprule
\textbf{Evaluation} & \textbf{Method} & \textbf{$N$} & \textbf{\apvr{} $\downarrow$} & \textbf{TLR $\downarrow$} & \textbf{\tsr{} $\uparrow$} & \textbf{ADF1 $\uparrow$} \\
\midrule
MCP server schemas & \textsc{NoProtection} & 26  & 0.842 & 0.923 & 1.000 & 0.271 \\
                   & \method{}             & 26  & \textbf{0.031} & \textbf{0.038} & \textbf{1.000} & \textbf{0.902} \\
\midrule
AgentDAM tasks     & \textsc{NoProtection} & 123 & 0.746 & 0.797 & \textbf{0.967} & 0.304 \\
                   & \method{}             & 123 & \textbf{0.128} & \textbf{0.203} & 0.935 & \textbf{0.768} \\
\bottomrule
\end{tabular}
\end{table}

\method{} substantially improves \apvr{}, TLR, and ADF1 in both external evaluations. It preserves MCP task success and keeps AgentDAM task success close to the unprotected agent (0.935 versus 0.967). \dataset{} provides controlled cross-condition evidence, while the MCP and AgentDAM evaluations test transfer to independently designed schemas, an independent benchmark, new domains, and a different agent interface. Broader generalization remains to be tested.

\subsection{Ablation of Disclosure-Justification Signals}
\label{sec:ablation}

\begin{figure}[t]
\centering
\includegraphics[width=0.98\linewidth]{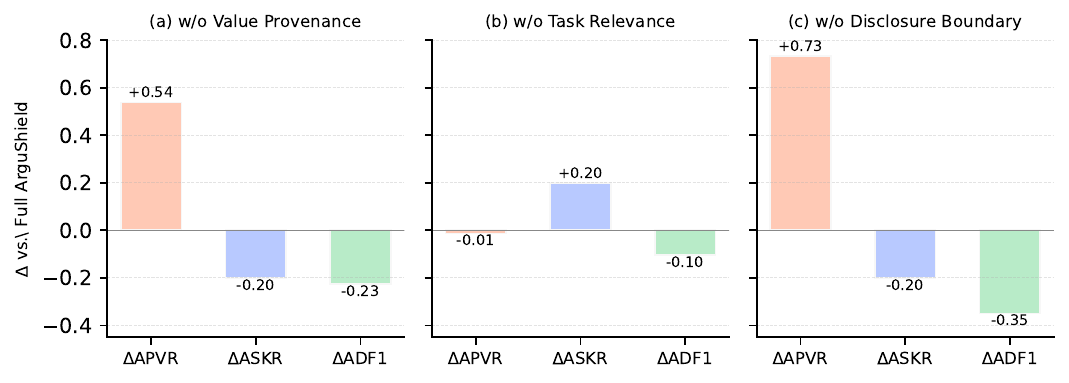}
\caption{Signal ablation.}
\label{fig:ablation}
\end{figure}

The ablation removes one disclosure-justification signal at a time and reports changes relative to full \method{}, macro-averaged over models and splits (Figure~\ref{fig:ablation}). We focus on \apvr{} for unauthorized silent release, \askr{} for confirmation behavior, and ADF1 for action-level alignment.

Removing the provenance signal sharply increases \apvr{} and lowers ADF1, showing that source grounding is a core privacy signal; the lower \askr{} indicates that some confirmation-needed values become silent releases. Removing Task Relevance produces little \apvr{} change but increases \askr{} and lowers ADF1, suggesting weaker precision in distinguishing task-needed from merely plausible values. Removing Disclosure Boundary yields the largest \apvr{} increase and ADF1 drop, confirming that receiver-specific restrictions are essential. Overall, provenance and boundary checks mainly prevent unauthorized silent release, while Task Relevance improves \keep{}/\ask{}/\hide{} assignment.

\subsection{Diagnostic Analysis}
\label{sec:diagnostic-analysis}

\begin{figure}[t]
\centering
\includegraphics[width=0.98\linewidth]{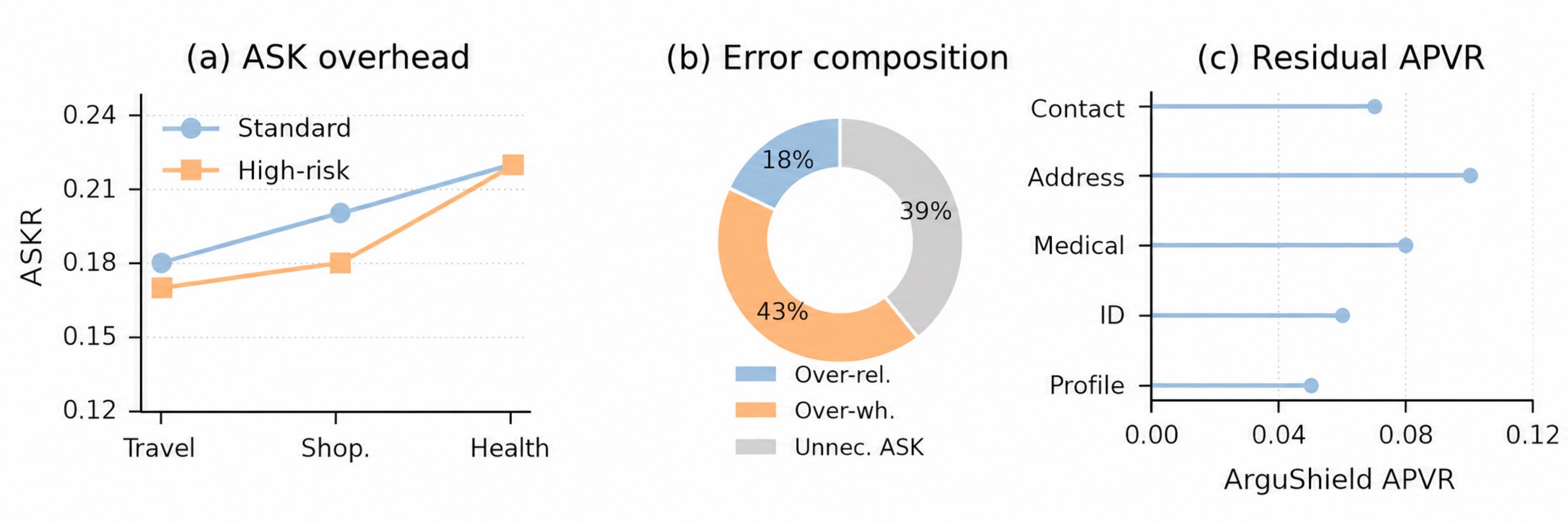}
\caption{Diagnostic analysis.}
\label{fig:diagnostic}
\end{figure}

We further inspect whether the privacy gains come from excessive confirmation, conservative withholding, or more precise release decisions. Figure~\ref{fig:diagnostic} summarizes confirmation overhead, residual action-level errors, and residual disclosure by value type.

Confirmation remains targeted rather than whole-call based. On completed tasks, Travel, Shopping, and Healthcare have Standard/High-risk \askr{} values of 0.18/0.17, 0.20/0.18, and 0.22/0.22, yielding overall rates of 0.20 and 0.19. Residual errors comprise over-release (18\%), over-withholding (43\%), and unnecessary confirmation (39\%). Residual \apvr{} remains low across Contact, Address, Medical, ID, and Preference/Profile values (0.07, 0.10, 0.08, 0.06, and 0.05).

\subsection{Disclosure Frequency and Consequence}
\label{sec:consequence-analysis}

\apvr{} intentionally measures violation frequency rather than disclosure consequence. The Standard and High-risk splits differ in receiver, purpose, and value-release constraints, so their aggregate \apvr{} values should not be interpreted as a ranking of disclosure severity. We therefore add an orthogonal consequence-tier analysis. Both splits contain Low, Moderate, and High consequence values, and we recompute \apvr{} within each tier using only the corresponding gold \ask{}/\hide{} opportunities.

\begin{table}[t]
\centering
\footnotesize
\renewcommand{\arraystretch}{1.15}
\setlength{\tabcolsep}{4.0pt}
\caption{Residual \method{} violations stratified by disclosure consequence. Consequence tiers complement rather than replace the frequency-based \apvr{} metric.}
\label{tab:consequence-tier}
\begin{tabular}{p{0.12\linewidth}p{0.36\linewidth}>{\centering\arraybackslash}p{0.17\linewidth}>{\centering\arraybackslash}p{0.17\linewidth}}
\toprule
\textbf{Tier} & \textbf{Typical values} & \textbf{Standard \apvr{} $\downarrow$} & \textbf{High-risk \apvr{} $\downarrow$} \\
\midrule
Low      & Preferences, scheduling, and contextual details                  & 0.552 & 0.420 \\
Moderate & Contact, address, and profile/account identifiers                & 0.075 & 0.058 \\
High     & Medical, financial, legal, credentials, and government IDs       & 0.020 & 0.009 \\
\bottomrule
\end{tabular}
\end{table}

Residual violation rates decrease from Low to Moderate to High consequence on both splits (Table~\ref{tab:consequence-tier}). We therefore retain \apvr{} as the primary measure of violation frequency and use consequence tiers as a complementary diagnostic of which disclosures remain after mediation.

\subsection{Confirmation-Policy Sensitivity}
\label{sec:confirmation-policy}

The main evaluation treats task-required \ask{} values as recoverable under simulated user confirmation. To quantify sensitivity to this assumption, we compare three deterministic confirmation policies. \textsc{Approve-all} approves every \ask{} value; \textsc{Bounded-k1} approves at most the first \ask{} value in deterministic flattened-argument order; and \textsc{Reject-all} approves none. Approved-Ask Rate (AAR) is the fraction of all candidate values that are assigned \ask{} and subsequently approved.

\begin{table}[t]
\centering
\small
\renewcommand{\arraystretch}{1.12}
\setlength{\tabcolsep}{5.0pt}
\caption{Sensitivity to simulated confirmation policy.}
\label{tab:confirmation-policy}
\begin{tabular}{lcccc}
\toprule
\textbf{Policy} & \textbf{Standard \tsr{} $\uparrow$} & \textbf{Standard AAR} & \textbf{High-risk \tsr{} $\uparrow$} & \textbf{High-risk AAR} \\
\midrule
\textsc{Approve-all} & 0.971 & 0.194 & 0.967 & 0.188 \\
\textsc{Bounded-k1}  & 0.971 & 0.114 & 0.967 & 0.109 \\
\textsc{Reject-all}  & 0.929 & 0.000 & 0.933 & 0.000 \\
\bottomrule
\end{tabular}
\end{table}

\textsc{Bounded-k1} reduces approved \ask{} values by approximately 41\% relative to \textsc{Approve-all} on both splits while preserving the same \tsr{}. \textsc{Reject-all} releases no \ask{} values, and \tsr{} decreases to 0.929 on Standard and 0.933 on High-risk tasks. \apvr{} is unchanged across confirmation policies because it measures unauthorized silent release rather than disclosure following simulated approval. This analysis quantifies a controllable privacy--interaction--utility trade-off under simulated confirmation. It does not replace a real-user confirmation study.

%% file: sections/limitations.tex
\section{Limitations}
\label{sec:limitations}

\method{} mediates release at the third-party tool boundary before execution and cannot retract information after release. Final responses, shared memory, inter-agent messages, and multi-tool orchestration are outside \method{}'s current scope. Related work identifies privacy risks from speculative tool dispatch \citep{mohammadi2026ghost} and sensitive inferences formed by combining tool outputs \citep{qiao2025topr}. \method{} is therefore one enforceable layer for structured tool inputs, not a complete privacy architecture.

Release decisions also depend on evidence quality. Ambiguous intent, underspecified schemas, incomplete context, or domain-specific norms can still produce conservative deferral or incorrect release despite auditable checks over Value Provenance, Task Relevance, and Disclosure Boundary. \dataset{} is similarly diagnostic: its controlled values and stable schemas make over-disclosure measurable, but do not capture the diversity of deployment traces, organizational policies, or legal regimes \citep{zharmagambetov2025agentdam,mireshghallah2024confaide,li2025privaci}.

Finally, we evaluate recoverability under simulated confirmation for \ask{} values rather than live human-in-the-loop interaction. Real users may reject or misunderstand confirmation requests, and excessive prompting can add interaction cost. Future work should study confirmation design and adaptive policies that preserve minimum-necessary disclosure with lower overhead.

%% file: sections/conclusion.tex
\section{Conclusion}
\label{sec:conclusion}

Tool-calling agents can complete user tasks while still exposing values that the current task or receiving tool does not justify. We formalize this failure as \emph{tool-argument over-disclosure}, a violation of \emph{minimum-necessary disclosure} at the third-party tool boundary.

\method{} addresses this risk through pre-execution release mediation. It treats a model-proposed invocation as an executable draft, evaluates candidate values using Value Provenance, Task Relevance, and Disclosure Boundary, and assigns each value to \keep{}, \ask{}, or \hide{} before execution.

We evaluate \method{} on \dataset{}, a 120-task benchmark across Travel, Shopping, and Healthcare. Experiments show that unprotected agents frequently release unnecessary private values despite high task success, while prompt-only protection reduces unauthorized disclosure at a substantial cost to task success. \method{} substantially reduces unauthorized silent release on both Standard and High-risk tasks while preserving recoverable task success under simulated user confirmation. These results support value-level mediation as a complementary enforcement layer at the tool-execution boundary.

%% file: main.bbl
\begin{thebibliography}{35}
\providecommand{\natexlab}[1]{#1}
\providecommand{\url}[1]{\texttt{#1}}
\expandafter\ifx\csname urlstyle\endcsname\relax
  \providecommand{\doi}[1]{doi: #1}\else
  \providecommand{\doi}{doi: \begingroup \urlstyle{rm}\Url}\fi

\bibitem[Alizadeh et~al.(2025)Alizadeh, Samei, Stetsenko, and
  Gilardi]{alizadeh2025promptleakage}
Meysam Alizadeh, Zeynab Samei, Daria Stetsenko, and Fabrizio Gilardi.
\newblock Simple prompt injection attacks can leak personal data observed by
  {LLM} agents during task execution, 2025.
\newblock URL \url{https://arxiv.org/abs/2506.01055}.

\bibitem[Bagdasarian et~al.(2024)Bagdasarian, Yi, Ghalebikesabi, Kairouz,
  Gruteser, Oh, Balle, and Ramage]{bagdasarian2024airgapagent}
Eugene Bagdasarian, Ren Yi, Sahra Ghalebikesabi, Peter Kairouz, Marco Gruteser,
  Sewoong Oh, Borja Balle, and Daniel Ramage.
\newblock {AirGapAgent}: Protecting privacy-conscious conversational agents.
\newblock In \emph{Proceedings of the 2024 on ACM SIGSAC Conference on Computer
  and Communications Security}, CCS '24, page 3868–3882, New York, NY, USA,
  2024. Association for Computing Machinery.
\newblock ISBN 9798400706363.
\newblock \doi{10.1145/3658644.3690350}.
\newblock URL \url{https://doi.org/10.1145/3658644.3690350}.

\bibitem[Costa et~al.(2025)Costa, Köpf, Kolluri, Paverd, Russinovich, Salem,
  Tople, Wutschitz, and Zanella-Béguelin]{costa2025fides}
Manuel Costa, Boris Köpf, Aashish Kolluri, Andrew Paverd, Mark Russinovich,
  Ahmed Salem, Shruti Tople, Lukas Wutschitz, and Santiago Zanella-Béguelin.
\newblock Securing ai agents with information-flow control, 2025.
\newblock URL \url{https://arxiv.org/abs/2505.23643}.

\bibitem[Debenedetti et~al.(2024)Debenedetti, Zhang, Balunovic, Beurer-Kellner,
  Fischer, and Tram{\`e}r]{debenedetti2024agentdojo}
Edoardo Debenedetti, Jie Zhang, Mislav Balunovic, Luca Beurer-Kellner, Marc
  Fischer, and Florian Tram{\`e}r.
\newblock {AgentDojo}: A dynamic environment to evaluate prompt injection
  attacks and defenses for {LLM} agents.
\newblock In \emph{The Thirty-eighth Conference on Neural Information
  Processing Systems Datasets and Benchmarks Track}, 2024.
\newblock URL \url{https://openreview.net/forum?id=m1YYAQjO3w}.

\bibitem[Debenedetti et~al.(2025)Debenedetti, Shumailov, Fan, Hayes, Carlini,
  Fabian, Kern, Shi, Terzis, and Tramèr]{debenedetti2025camel}
Edoardo Debenedetti, Ilia Shumailov, Tianqi Fan, Jamie Hayes, Nicholas Carlini,
  Daniel Fabian, Christoph Kern, Chongyang Shi, Andreas Terzis, and Florian
  Tramèr.
\newblock Defeating prompt injections by design, 2025.
\newblock URL \url{https://arxiv.org/abs/2503.18813}.

\bibitem[Fan et~al.(2026)Fan, Li, Tian, Wang, Li, and Wang]{fan2026pact}
Linfeng Fan, Ziwei Li, Yuan Tian, Yichen Wang, Rongsheng Li, and Xiong Wang.
\newblock The granularity mismatch in agent security: Argument-level provenance
  solves enforcement and isolates the {LLM} reasoning bottleneck, 2026.
\newblock URL \url{https://arxiv.org/abs/2605.11039}.

\bibitem[Hu et~al.(2026)Hu, Liu, Meng, and Zhao]{hu2026toolprivacybench}
Shijing Hu, Liang Liu, Zhu Meng, and Zhicheng Zhao.
\newblock Toolprivacybench: Benchmarking purpose-bound privacy in tool-using
  {LLM} agents, 2026.
\newblock URL \url{https://arxiv.org/abs/2606.28061}.

\bibitem[Kan et~al.(2023)Kan, Qiao, Yu, Peng, Gao, and
  Li]{kan2023privacytextsanitization}
Zhigang Kan, Linbo Qiao, Hao Yu, Liwen Peng, Yifu Gao, and Dongsheng Li.
\newblock Protecting user privacy in remote conversational systems: A
  privacy-preserving framework based on text sanitization, 2023.
\newblock URL \url{https://arxiv.org/abs/2306.08223}.

\bibitem[Li et~al.(2025)Li, Hu, Jing, Chen, Hu, Han, Chu, Hu, and
  Song]{li2025privaci}
Haoran Li, Wenbin Hu, Huihao Jing, Yulin Chen, Qi~Hu, Sirui Han, Tianshu Chu,
  Peizhao Hu, and Yangqiu Song.
\newblock {P}riva{CI}-bench: Evaluating privacy with contextual integrity and
  legal compliance.
\newblock In Wanxiang Che, Joyce Nabende, Ekaterina Shutova, and Mohammad~Taher
  Pilehvar, editors, \emph{Proceedings of the 63rd Annual Meeting of the
  Association for Computational Linguistics (Volume 1: Long Papers)}, pages
  10544--10559, Vienna, Austria, July 2025. Association for Computational
  Linguistics.
\newblock ISBN 979-8-89176-251-0.
\newblock \doi{10.18653/v1/2025.acl-long.518}.
\newblock URL \url{https://aclanthology.org/2025.acl-long.518/}.

\bibitem[Lin et~al.(2026)Lin, Li, Pan, Shi, Yao, and Xu]{lin2026sapa}
Zhixin Lin, Jungang Li, Shidong Pan, Yibo Shi, Yue Yao, and Dongliang Xu.
\newblock Mind the third eye! benchmarking privacy awareness in {MLLM}-powered
  smartphone agents.
\newblock In \emph{The Fortieth Annual AAAI Conference on Artificial
  Intelligence}, 2026.
\newblock \doi{10.1609/aaai.v40i42.40874}.
\newblock URL \url{https://ojs.aaai.org/index.php/AAAI/article/view/40874}.

\bibitem[Liu et~al.(2024)Liu, Hoang, Zhang, Zhu, Lan, Kokane, Tan, Yao, Liu,
  Feng, N, Yang, Savarese, Niebles, Wang, Heinecke, and Xiong]{liu2024apigen}
Zuxin Liu, Thai~Quoc Hoang, Jianguo Zhang, Ming Zhu, Tian Lan, Shirley Kokane,
  Juntao Tan, Weiran Yao, Zhiwei Liu, Yihao Feng, Rithesh~R N, Liangwei Yang,
  Silvio Savarese, Juan~Carlos Niebles, Huan Wang, Shelby Heinecke, and Caiming
  Xiong.
\newblock {APIGen}: Automated {PI}peline for generating verifiable and diverse
  function-calling datasets.
\newblock In \emph{The Thirty-eighth Conference on Neural Information
  Processing Systems Datasets and Benchmarks Track}, 2024.
\newblock URL \url{https://openreview.net/forum?id=Jfg3vw2bjx}.

\bibitem[Lu et~al.(2025)Lu, Holleis, Zhang, Aumayer, Nan, Bai, Ma, Ma, Li, Yin,
  Wang, and Pang]{lu2025toolsandbox}
Jiarui Lu, Thomas Holleis, Yizhe Zhang, Bernhard Aumayer, Feng Nan, Haoping
  Bai, Shuang Ma, Shen Ma, Mengyu Li, Guoli Yin, Zirui Wang, and Ruoming Pang.
\newblock {T}ool{S}andbox: A stateful, conversational, interactive evaluation
  benchmark for {LLM} tool use capabilities.
\newblock In Luis Chiruzzo, Alan Ritter, and Lu~Wang, editors, \emph{Findings
  of the Association for Computational Linguistics: NAACL 2025}, pages
  1160--1183, Albuquerque, New Mexico, April 2025. Association for
  Computational Linguistics.
\newblock ISBN 979-8-89176-195-7.
\newblock \doi{10.18653/v1/2025.findings-naacl.65}.
\newblock URL \url{https://aclanthology.org/2025.findings-naacl.65/}.

\bibitem[Mireshghallah et~al.(2024)Mireshghallah, Kim, Zhou, Tsvetkov, Sap,
  Shokri, and Choi]{mireshghallah2024confaide}
Niloofar Mireshghallah, Hyunwoo Kim, Xuhui Zhou, Yulia Tsvetkov, Maarten Sap,
  Reza Shokri, and Yejin Choi.
\newblock Can {LLM}s keep a secret? testing privacy implications of language
  models via contextual integrity theory.
\newblock In \emph{The Twelfth International Conference on Learning
  Representations}, 2024.
\newblock URL \url{https://openreview.net/forum?id=gmg7t8b4s0}.

\bibitem[Mohammadi et~al.(2026)Mohammadi, Klein, Arora, and
  Bindschaedler]{mohammadi2026ghost}
Bardia Mohammadi, Lars Klein, Akhil Arora, and Laurent Bindschaedler.
\newblock Ghost tool calls: Issue-time privacy for speculative agent tools,
  2026.
\newblock URL \url{https://arxiv.org/abs/2606.02483}.

\bibitem[Patil et~al.(2024)Patil, Zhang, Wang, and Gonzalez]{patil2024gorilla}
Shishir~G. Patil, Tianjun Zhang, Xin Wang, and Joseph~E. Gonzalez.
\newblock Gorilla: Large language model connected with massive apis.
\newblock In A.~Globerson, L.~Mackey, D.~Belgrave, A.~Fan, U.~Paquet,
  J.~Tomczak, and C.~Zhang, editors, \emph{Advances in Neural Information
  Processing Systems}, volume~37, pages 126544--126565. Curran Associates,
  Inc., 2024.
\newblock \doi{10.52202/079017-4020}.
\newblock URL
  \url{https://proceedings.neurips.cc/paper_files/paper/2024/file/e4c61f578ff07830f5c37378dd3ecb0d-Paper-Conference.pdf}.

\bibitem[Patil et~al.(2025)Patil, Mao, Yan, Ji, Suresh, Stoica, and
  Gonzalez]{patil2025bfcl}
Shishir~G Patil, Huanzhi Mao, Fanjia Yan, Charlie Cheng-Jie Ji, Vishnu Suresh,
  Ion Stoica, and Joseph~E. Gonzalez.
\newblock The berkeley function calling leaderboard ({BFCL}): From tool use to
  agentic evaluation of large language models.
\newblock In Aarti Singh, Maryam Fazel, Daniel Hsu, Simon Lacoste-Julien, Felix
  Berkenkamp, Tegan Maharaj, Kiri Wagstaff, and Jerry Zhu, editors,
  \emph{Proceedings of the 42nd International Conference on Machine Learning},
  volume 267 of \emph{Proceedings of Machine Learning Research}, pages
  48371--48392. PMLR, 13--19 Jul 2025.
\newblock URL \url{https://proceedings.mlr.press/v267/patil25a.html}.

\bibitem[Qiao et~al.(2026)Qiao, Liu, Yang, Zhou, and Hu]{qiao2025topr}
Yuxuan Qiao, Dongqin Liu, Hongchang Yang, Wei Zhou, and Songlin Hu.
\newblock Agent tools orchestration leaks more: Dataset, benchmark, and
  mitigation, 2026.
\newblock URL \url{https://arxiv.org/abs/2512.16310}.

\bibitem[Qin et~al.(2024)Qin, Liang, Ye, Zhu, Yan, Lu, Lin, Cong, Tang, Qian,
  Zhao, Hong, Tian, Xie, Zhou, Gerstein, dahai li, Liu, and
  Sun]{qin2024toolllm}
Yujia Qin, Shihao Liang, Yining Ye, Kunlun Zhu, Lan Yan, Yaxi Lu, Yankai Lin,
  Xin Cong, Xiangru Tang, Bill Qian, Sihan Zhao, Lauren Hong, Runchu Tian,
  Ruobing Xie, Jie Zhou, Mark Gerstein, dahai li, Zhiyuan Liu, and Maosong Sun.
\newblock Tool{LLM}: Facilitating large language models to master 16000+
  real-world {API}s.
\newblock In \emph{The Twelfth International Conference on Learning
  Representations}, 2024.
\newblock URL \url{https://openreview.net/forum?id=dHng2O0Jjr}.

\bibitem[Roh et~al.(2026)Roh, Bagdasarian, Haddadi, and
  Shamsabadi]{roh2026spillage}
Jaechul Roh, Eugene Bagdasarian, Hamed Haddadi, and Ali~Shahin Shamsabadi.
\newblock {SPILLage}: Agentic oversharing on the web, 2026.
\newblock URL \url{https://arxiv.org/abs/2602.13516}.

\bibitem[Ruan et~al.(2024)Ruan, Dong, Wang, Pitis, Zhou, Ba, Dubois, Maddison,
  and Hashimoto]{ruan2024toolemu}
Yangjun Ruan, Honghua Dong, Andrew Wang, Silviu Pitis, Yongchao Zhou, Jimmy Ba,
  Yann Dubois, Chris~J. Maddison, and Tatsunori Hashimoto.
\newblock Identifying the risks of {LM} agents with an {LM}-emulated sandbox.
\newblock In \emph{The Twelfth International Conference on Learning
  Representations}, 2024.
\newblock URL \url{https://openreview.net/forum?id=GEcwtMk1uA}.

\bibitem[Schick et~al.(2023)Schick, Dwivedi-Yu, Dessi, Raileanu, Lomeli,
  Hambro, Zettlemoyer, Cancedda, and Scialom]{schick2023toolformer}
Timo Schick, Jane Dwivedi-Yu, Roberto Dessi, Roberta Raileanu, Maria Lomeli,
  Eric Hambro, Luke Zettlemoyer, Nicola Cancedda, and Thomas Scialom.
\newblock Toolformer: Language models can teach themselves to use tools.
\newblock In A.~Oh, T.~Naumann, A.~Globerson, K.~Saenko, M.~Hardt, and
  S.~Levine, editors, \emph{Advances in Neural Information Processing Systems},
  volume~36, pages 68539--68551. Curran Associates, Inc., 2023.
\newblock URL
  \url{https://proceedings.neurips.cc/paper_files/paper/2023/file/d842425e4bf79ba039352da0f658a906-Paper-Conference.pdf}.

\bibitem[Shao et~al.(2024)Shao, Li, Shi, Liu, and Yang]{shao2024privacylens}
Yijia Shao, Tianshi Li, Weiyan Shi, Yanchen Liu, and Diyi Yang.
\newblock {PrivacyLens}: Evaluating privacy norm awareness of language models
  in action.
\newblock In \emph{The Thirty-eighth Conference on Neural Information
  Processing Systems Datasets and Benchmarks Track}, 2024.
\newblock URL \url{https://openreview.net/forum?id=CxNXoMnCKc}.

\bibitem[Shayesteh and Wilson(2026)]{shayesteh-wilson-2026-conventional}
Shahriar Shayesteh and Shomir Wilson.
\newblock From conventional web privacy to agentic disclosure: How tool schemas
  may invite {LLM} oversharing.
\newblock In Ivan Habernal, Sepideh Ghanavati, Sara Haghighi, Krithika Ramesh,
  Timour Igamberdiev, and Shomir Wilson, editors, \emph{Proceedings of the
  Seventh Workshop on Privacy in Natural Language Processing}, pages 1--6, San
  Diego, California, July 2026. Association for Computational Linguistics.
\newblock ISBN 979-8-89176-397-5.
\newblock \doi{10.18653/v1/2026.privatenlp-main.1}.
\newblock URL \url{https://aclanthology.org/2026.privatenlp-main.1/}.

\bibitem[Trivedi et~al.(2024)Trivedi, Khot, Hartmann, Manku, Dong, Li, Gupta,
  Sabharwal, and Balasubramanian]{trivedi2024appworld}
Harsh Trivedi, Tushar Khot, Mareike Hartmann, Ruskin Manku, Vinty Dong, Edward
  Li, Shashank Gupta, Ashish Sabharwal, and Niranjan Balasubramanian.
\newblock {A}pp{W}orld: A controllable world of apps and people for
  benchmarking interactive coding agents.
\newblock In Lun-Wei Ku, Andre Martins, and Vivek Srikumar, editors,
  \emph{Proceedings of the 62nd Annual Meeting of the Association for
  Computational Linguistics (Volume 1: Long Papers)}, pages 16022--16076,
  Bangkok, Thailand, August 2024. Association for Computational Linguistics.
\newblock \doi{10.18653/v1/2024.acl-long.850}.
\newblock URL \url{https://aclanthology.org/2024.acl-long.850/}.

\bibitem[Wang et~al.(2025)Wang, Zhou, Wang, Mo, Zhang, Lin, Jin, Wen, Zhang,
  Peng, and Wang]{wang2024hammerbench}
Jun Wang, Jiamu Zhou, Xihuai Wang, Xiaoyun Mo, Haoyu Zhang, Qiqiang Lin, Cheng
  Jin, Muning Wen, Weinan Zhang, Qiuying Peng, and Jun Wang.
\newblock {H}ammer{B}ench: Fine-grained function-calling evaluation in real
  mobile assistant scenarios.
\newblock In Wanxiang Che, Joyce Nabende, Ekaterina Shutova, and Mohammad~Taher
  Pilehvar, editors, \emph{Findings of the Association for Computational
  Linguistics: ACL 2025}, pages 3350--3376, Vienna, Austria, July 2025.
  Association for Computational Linguistics.
\newblock ISBN 979-8-89176-256-5.
\newblock \doi{10.18653/v1/2025.findings-acl.175}.
\newblock URL \url{https://aclanthology.org/2025.findings-acl.175/}.

\bibitem[Wang et~al.(2024)Wang, Xu, Ye, Yan, Shen, Zhang, Huang, and
  Sang]{wang2024mobileagent}
Junyang Wang, Haiyang Xu, Jiabo Ye, Ming Yan, Weizhou Shen, Ji~Zhang, Fei
  Huang, and Jitao Sang.
\newblock Mobile-agent: Autonomous multi-modal mobile device agent with visual
  perception.
\newblock In \emph{ICLR 2024 Workshop on Large Language Model (LLM) Agents},
  2024.
\newblock URL \url{https://openreview.net/forum?id=jE6pDYCnVF}.

\bibitem[Wang et~al.(2026)Wang, Ba, Liu, Pan, Wei, Su, Luan, and
  Du]{wang2026openclawsecurity}
Yuntao Wang, Jianle Ba, Han Liu, Yanghe Pan, Jintao Wei, Zhou Su, Tom~H. Luan,
  and Linkang Du.
\newblock Security of {OpenClaw} agents: Fundamentals, attacks, and
  countermeasures, 2026.
\newblock URL \url{https://arxiv.org/abs/2605.25435}.

\bibitem[Wen et~al.(2026)Wen, Zhang, Lian, Yi, Xie, and
  Yang]{wen2026contextualized}
Yule Wen, Yanzhe Zhang, Jianxun Lian, Xiaoyuan Yi, Xing Xie, and Diyi Yang.
\newblock Contextualized privacy defense for {LLM} agents.
\newblock In \emph{Forty-third International Conference on Machine Learning},
  2026.
\newblock URL \url{https://openreview.net/forum?id=b4QtBbSV3X}.

\bibitem[Xie et~al.(2024)Xie, Zhang, Chen, Li, Zhao, Cao, Hua, Cheng, Shin,
  Lei, Liu, Xu, Zhou, Savarese, Xiong, Zhong, and Yu]{xie2024osworld}
Tianbao Xie, Danyang Zhang, Jixuan Chen, Xiaochuan Li, Siheng Zhao, Ruisheng
  Cao, Toh~Jing Hua, Zhoujun Cheng, Dongchan Shin, Fangyu Lei, Yitao Liu,
  Yiheng Xu, Shuyan Zhou, Silvio Savarese, Caiming Xiong, Victor Zhong, and Tao
  Yu.
\newblock {OSWorld}: Benchmarking multimodal agents for open-ended tasks in
  real computer environments.
\newblock In A.~Globerson, L.~Mackey, D.~Belgrave, A.~Fan, U.~Paquet,
  J.~Tomczak, and C.~Zhang, editors, \emph{Advances in Neural Information
  Processing Systems}, volume~37, pages 52040--52094. Curran Associates, Inc.,
  2024.
\newblock \doi{10.52202/079017-1650}.
\newblock URL
  \url{https://proceedings.neurips.cc/paper_files/paper/2024/file/5d413e48f84dc61244b6be550f1cd8f5-Paper-Datasets_and_Benchmarks_Track.pdf}.

\bibitem[Yao et~al.(2023)Yao, Zhao, Yu, Du, Shafran, Narasimhan, and
  Cao]{yao2023react}
Shunyu Yao, Jeffrey Zhao, Dian Yu, Nan Du, Izhak Shafran, Karthik Narasimhan,
  and Yuan Cao.
\newblock {ReAct}: Synergizing reasoning and acting in language models.
\newblock In \emph{International Conference on Learning Representations
  (ICLR)}, 2023.

\bibitem[Yao et~al.(2025)Yao, Shinn, Razavi, and Narasimhan]{yao2025taubench}
Shunyu Yao, Noah Shinn, Pedram Razavi, and Karthik~R. Narasimhan.
\newblock $\tau$-bench: A benchmark for tool-agent-user interaction in
  real-world domains.
\newblock In \emph{The Thirteenth International Conference on Learning
  Representations}, 2025.
\newblock URL \url{https://openreview.net/forum?id=roNSXZpUDN}.

\bibitem[Zhang et~al.(2025)Zhang, Yang, Liu, Li, Han, Chen, Huang, Fu, and
  Yu]{zhang2025appagent}
Chi Zhang, Zhao Yang, Jiaxuan Liu, Yanda Li, Yucheng Han, Xin Chen, Zebiao
  Huang, Bin Fu, and Gang Yu.
\newblock {AppAgent}: Multimodal agents as smartphone users.
\newblock In \emph{Proceedings of the 2025 CHI Conference on Human Factors in
  Computing Systems}, CHI '25, New York, NY, USA, 2025. Association for
  Computing Machinery.
\newblock ISBN 9798400713941.
\newblock \doi{10.1145/3706598.3713600}.
\newblock URL \url{https://doi.org/10.1145/3706598.3713600}.

\bibitem[Zhang et~al.(2024)Zhang, Xu, Ba, Wang, Hong, Liu, Qin, and
  Ren]{zhang2024privacyasst}
Xinyu Zhang, Huiyu Xu, Zhongjie Ba, Zhibo Wang, Yuan Hong, Jian Liu, Zhan Qin,
  and Kui Ren.
\newblock {PrivacyAsst}: Safeguarding user privacy in tool-using large language
  model agents.
\newblock \emph{IEEE Transactions on Dependable and Secure Computing},
  21\penalty0 (6):\penalty0 5242--5258, 2024.
\newblock \doi{10.1109/TDSC.2024.3372777}.

\bibitem[Zharmagambetov et~al.(2025)Zharmagambetov, Guo, Evtimov, Pavlova,
  Salakhutdinov, and Chaudhuri]{zharmagambetov2025agentdam}
Arman Zharmagambetov, Chuan Guo, Ivan Evtimov, Maya Pavlova, Ruslan
  Salakhutdinov, and Kamalika Chaudhuri.
\newblock {AgentDAM}: Privacy leakage evaluation for autonomous web agents.
\newblock In \emph{Advances in Neural Information Processing Systems 38,
  Datasets and Benchmarks Track}, 2025.
\newblock URL
  \url{https://proceedings.neurips.cc/paper_files/paper/2025/hash/c9826b9ea5e1b49b256329934a578d83-Abstract-Datasets_and_Benchmarks_Track.html}.

\bibitem[Zhou et~al.(2024)Zhou, Xu, Zhu, Zhou, Lo, Sridhar, Cheng, Ou, Bisk,
  Fried, Alon, and Neubig]{zhou2024webarena}
Shuyan Zhou, Frank~F Xu, Hao Zhu, Xuhui Zhou, Robert Lo, Abishek Sridhar,
  Xianyi Cheng, Tianyue Ou, Yonatan Bisk, Daniel Fried, Uri Alon, and Graham
  Neubig.
\newblock {WebArena}: A realistic web environment for building autonomous
  agents.
\newblock In B.~Kim, Y.~Yue, S.~Chaudhuri, K.~Fragkiadaki, M.~Khan, and Y.~Sun,
  editors, \emph{International Conference on Learning Representations}, volume
  2024, pages 15585--15606, 2024.
\newblock URL
  \url{https://proceedings.iclr.cc/paper_files/paper/2024/file/4410c0711e9154a7a2d26f9b3816d1ef-Paper-Conference.pdf}.

\end{thebibliography}
